# Long Exciton Dephasing Time and Coherent Phonon Coupling in CsPbBr$_2$Cl Perovskite Nanocrystals

*Michael A. Becker[1,2], Lorenzo Scarpelli[3], Georgian Nedelcu[4,5], Gabriele Rainò[4,5], Francesco Masia[3], Paola Borri[3,6], Thilo Stöferle[1], Maksym V. Kovalenko[4,5]\*, Wolfgang Langbein[3]\*, Rainer F. Mahrt[1]\**

AUTHOR ADDRESS

[1]IBM Research−Zurich, Säumerstrasse 4, 8803 Rüschlikon, Switzerland
[2]Optical Materials Engineering Laboratory, ETH Zürich, 8092 Zürich, Switzerland.
[3]School of Physics and Astronomy, Cardiff University, The Parade, Cardiff CF243AA, United Kingdom
[4]Institute of Inorganic Chemistry, Department of Chemistry and Applied Bioscience, ETH Zürich, 8093 Zürich, Switzerland.
[5]Laboratory of Thin Films and Photovoltaics, Empa – Swiss Federal Laboratories for Materials Science and Technology, 8600 Dübendorf, Switzerland.
[6]Cardiff University School of Biosciences, Museum Avenue, Cardiff CF10 3AX, United Kingdom

ABSTRACT

Fully-inorganic cesium lead halide perovskite nanocrystals (NCs) have shown to exhibit outstanding optical properties such as wide spectral tunability, high quantum yield, high oscillator strength as well as blinking-free single photon emission and low spectral diffusion. Here, we report measurements of the coherent and incoherent exciton dynamics on the 100 fs to 10 ns timescale, determining dephasing and density decay rates in these NCs. The experiments are performed on $CsPbBr_2Cl$ NCs using transient resonant three-pulse four-wave mixing (FWM) in heterodyne detection at temperatures ranging from 5 K to 50 K. We found a low-temperature exciton dephasing time of 24.5±1.0 ps, inferred from the decay of the photon-echo amplitude at 5 K, corresponding to a homogeneous linewidth (FWHM) of 54±5 μeV. Furthermore, oscillations in the photon-echo signal on a picosecond timescale are observed and attributed to coherent coupling of the exciton to a quantized phonon mode with 3.45 meV energy.

TEXT

Recently, a new type of colloidal nanocrystals (NCs) has emerged for opto-electronic applications which combines simplicity in synthesis with great spectral flexibility and exceptional optical properties. Fully inorganic cesium lead halide perovskite NCs ($CsPbX_3$, where X = Cl, Br, I or mixture thereof)[1,2] have shown outstanding optical properties such as wide spectral tunability and high oscillator strength.[3] These NCs can be synthesized with precise compositional and size control, and show room-temperature photoluminescence (PL) quantum yields (QY) of 60 – 90% (ref. 1). Moreover, perovskite NCs have attracted a lot of interest due to their large absorption coefficient and gain for optically pumped lasing devices.[4] At cryogenic temperatures, the PL decay is mostly radiative with lifetimes in the few hundred picosecond range, depending on size and composition of the NCs. This results from a "giant oscillator strength" in the intermediate confinement regime with an exciton Bohr radius of 5 − 7 nm for cesium lead bromide-chloride ($CsPbBr_2Cl$) NCs and a bright lowest triplet state manifold[5]. Furthermore, almost blinking-free single photon emission and marginal spectral diffusion have been reported for $CsPbX_3$ quantum dots at low temperature.[3,6] These remarkable features make perovskite-type lead halide NCs a prime candidate for the observation of strong light-matter interaction, e.g., showing coherent cooperative emission[7] or creating exciton-polaritons by embedding them in high-finesse optical cavities, as shown for CVD-grown fully-inorganic lead halide perovskite nanowires[8,9] and nano-platelets.[10] However, the timescale of such coherent coupling is limited by the exciton dephasing, which is still unknown for this material.

Transient four-wave-mixing (TFWM) spectroscopy is a powerful method allowing to directly measure the loss of quantum coherence characterized by a dephasing time $T_2$. It has been applied on various materials such as ruby crystals[11], atoms[12], molecules[13,14] and semiconductor nanostructures.[15–17] In general, the dephasing of excitons in semiconductor materials is caused by elastic and inelastic scattering processes with phonons and charge carriers, and by radiative population decay.[18]

We studied the exciton dephasing and population dynamics using three-pulse degenerate TFWM spectroscopy on an ensemble of $CsPbBr_2Cl$ NCs. Previous measurements on colloidal CdSe-based NCs[19,20] and nanoplatelets[21] revealed a strong dependence of the $T_2$ time on the material and its shape and size. The investigated cubic $CsPbBr_2Cl$ NCs with edge lengths of $10 \pm 1$ nm were synthesized as discussed in the Supplementary Information (SI), and possess the 3D-perovskite orthorhombic crystal structure (*Pnma* space group), shown in Figure 1a. Single quantum dot (QD) spectroscopy at cryogenic temperatures revealed that the emission of individual CsPbBr$_2$Cl NCs exhibits a PL full-width at half-maximum (FWHM) below 1 meV, and a fine structure with an average energy splitting around 1 meV.[3,22] The exciton decay, measured at 5 K using non-resonant excitation, is mostly radiative with decay times of 180 – 250 ps. This is 1000 times faster than in CdSe/ZnS QDs[23] at cryogenic temperatures, and attributed to high oscillator strength due to larger exciton coherence volume, and the absence of a low-energy dark state.[5]

We performed TFWM experiments on films prepared by drop-casting a solution of NCs and polystyrene in toluene on *c*-cut quartz substrates (see SI). At room-temperature, the PL emission (see Figure 1c) is centered at a photon energy of 2.63 eV, and exhibits a Stokes shift of about 70 meV with respect to the ground-state exciton absorption resonance. At 5 K, the PL emission red-shifts to 2.54 eV, which is a known feature of lead-based semiconductor NCs like PbS and PbSe[24], and the PL FWHM decreases from 85 meV to 20 meV. The TFWM experiments have been performed by resonant excitation of the NCs at 2.6 eV with femtosecond pulses (120 fs intensity FWHM) from the second harmonic of a Ti:Sapphire oscillator with 76 MHz repetition rate (for details of the experimental setup see Naeem *et al.*, ref. 21). The first excitation pulse (P$_1$) with wavevector $\boldsymbol{k}_1$ induces a coherent polarization of the emitters in the inhomogeneous sample, which is then subject to dephasing. After a time delay $\tau_{12}$, a second pulse (P$_2$) converts the polarization into a population density grating. The third pulse (P$_3$), that arrives on the sample after a time delay $\tau_{23}$, is diffracted by the density grating, creating a FWM signal with a wavevector of $\boldsymbol{k}_S = \boldsymbol{k}_3 + \boldsymbol{k}_2 - \boldsymbol{k}_1$ (refs. 20,21). To investigate the exciton population dynamics, we set the time delay between the first and the second pulse to zero ($\tau_{12} = 0$ ps), and measure the FWM signal as a function of the time delay $\tau_{23}$. We use a spatial selection geometry to suppress the transmitted excitation pulses, and then further discriminate the FWM signal from the exciting pulses using a heterodyne technique, in which the pulse train $P_i$ is radio-frequency shifted by $\Omega_i$ ($i$=1, 2, 3), resulting in a frequency-shifted FWM field which is detected by its interference with a reference pulse.[21] In Figure 2a, the measured FWM field amplitude (black) and phase (blue) at 5 K with their respective fits are shown. We fit amplitude and phase with a bi-exponential response function to quantify the population dynamics with decay rates $\Gamma_1 > \Gamma_2$, as explained in the SI. Superimposed onto the population decay, we observe in the initial dynamics damped oscillations with a period of about $1.2 \pm 0.1$ ps, which we interpret as coherent phonon interactions, as we will discuss in more detail further below.

At 5 K, the FWM field amplitude decays with two distinct time constants. The fast decay time $\tau_1 = 28.2 \pm 0.8$ ps with a relative amplitude of $\frac{A_1}{A_1+A_2} = 0.89 \pm 0.07$ (see Figure S2 in the SI) corresponds to a decay rate $\Gamma_1 = \frac{1}{\tau_1} = 35.5 \pm 1.0$ ns$^{-1}$, i.e. a linewidth of $\hbar\Gamma_1 = 23.4 \pm 2$ µeV (Figure 2b). This rate is independent of excitation power (see Figure S3 in the SI), and we note that the decay rate is higher compared to non-resonantly excited PL.[5] A qualitatively similar difference between lifetimes observed in resonant FWM and non-resonant PL is observed in CdSe nanoplatelets[21] and CdSe/ZnS NCs[19], and is attributed to the resonant excitation. Because the lowest state in cesium lead halide NCs is bright[5], relaxation to the dark state is not relevant at low temperatures. With increasing temperatures the decay rate $\Gamma_1$ is decreasing, which is interpreted to arise from the occupation of the dark fine-structure state at $T = 20$ K, with the level structure proposed in ref. 5.

The second decay component is four orders of magnitude slower, $\Gamma_2 = 0.037 \pm 0.009$ ns$^{-1}$ at 5 K, which is below the repetition rate in the experiment, and has a low weight of about 10% of the first one. In contrast to the first decay component, $\Gamma_2$ is increasing with temperature, as shown in Figure 2c. We assign this component to trap or defect states present in a small fraction of the NCs, increasing their decay rate with temperature by thermal activation. The relative amplitude of the components is temperature independent within error (see SI). The highest excitation density is estimated to excite up to 0.08 excitons per excitation pulse per NC, ruling out significant multi-exciton effects.

The dephasing time can be extracted from the decay of the photon echo, which we measured using three-pulse FWM spectroscopy in a heterodyne detection scheme (see SI). We scan the time delay $\tau_{12}$ between the first and the second pulse, while choosing a positive time delay $\tau_{23} = 1$ ps to avoid instantaneous non-resonant non-linearities.[21] The photon echo is then emitted at time $\tau_{12}$ after the third pulse P$_3$, as depicted in the inset of Figure 3a. The time-integrated FWM field amplitude as a function of delay time $\tau_{12}$ is shown in Figure 3a for various temperatures. The FWM amplitude shows a bi-exponential behavior up to 41 K. For higher temperatures, it can be described with a fast mono-exponential decay. The initial fast decay of the amplitude proportional to $\exp(-2\gamma_1\tau_{12})$ in the time domain corresponds to a linewidth $2\hbar\gamma_1 = 4.37 \pm 0.16$ meV at 5 K. It is attributed to phonon-assisted transitions and a quantum beat of the fine-structure split states, which show a distribution of splittings in the meV energy range as they vary from NC to NC. Since the exciton-exciton interaction is significant between any of the states, the density gratings of the states are adding up, constructively interfering at $\tau_{12} = 0$, and the beat with a wide distribution of frequencies results in a decay over the timescale of the inverse splitting. Assuming we excite the three bright states uniformly, a decay of the signal by a factor of three over the timescale of about a picosecond, given by the inverse energy splitting, is expected. Additionally, phonon-assisted transitions will contribute, as observed in other 3D confined systems.[19,20,25] However, the large extension of the excitons inside the individual NCs, which is the origin of the giant oscillator strength, is reducing their weight in the signal. This is confirmed in low-temperature single NC PL spectra, which do not show significant phonon-assisted emission, with an estimated zero phonon line (ZPL) weight of 0.93 (see SI). At 5 K, we find a linewidth of $2\hbar\gamma_1 = 4.37 \pm 0.16$ meV, which increases slowly as a function of temperature (Figure 3b). In general, the homogeneous linewidth of each fine-structure transition in the spectral domain is composed of a broad acoustic phonon band that corresponds to the fast initial dephasing, which is superimposed on a sharp Lorentzian-shaped ZPL, corresponding to the long exponential dephasing in the time domain.[25] From the second decay component of the photon echo, we can therefore deduce the ZPL width $2\hbar\gamma_2 = 54 \pm 5$ µeV, corresponding to a dephasing time $T_2 = 24.5 \pm 1.0$ ps at 5 K. In PL measurements of single NCs at 5 K, linewidths of typically a few hundred µeV are found for CsPbBr$_2$Cl NCs[3] (see SI). The value obtained by FWM is consistent with this, considering that spectral diffusion is typically affecting single NC PL.[26] The temperature dependence of the homogeneous linewidth is plotted in Figure 3c. The solid red line is a temperature-activated fit of the ZPL width using $2\hbar\gamma_2 = \hbar\gamma_0 + \dfrac{b}{e^{\frac{\Delta}{k_BT}}-1}$, with the scattering energy $b = 0.23 \pm 0.09$ meV, and the zero-temperature extrapolated linewidth $\hbar\gamma_0 = 39.6 \pm 7.1$ µeV. The activation energy $\Delta = 1.23 \pm 0.35$ meV is consistent with the reported fine-structure splitting from single NC spectroscopy measurements[5] (see SI), suggesting that phonon-assisted scattering into different fine-structure states is responsible for the dephasing. In Figure 3d the temperature dependence of the nominal ZPL weight $Z = \sqrt[3]{\dfrac{C_2}{C_1+C_2}}$ is plotted, where $C_1$ and $C_2$ are the decay amplitudes of the fast and long decay component of the photon echo, respectively. The nominal ZPL weight is $Z = 0.44$ and remains constant up to 10 K and then starts to decrease with increasing temperature. As discussed above we attribute the initial decay of the photon echo mostly to an overdamped beat between the fine-structure states, so that the reported nominal ZPL weight is lower than the real ZPL weight of the individual bright transitions. For equal weights of three transitions in the fine-structure beat, the amplitude will decay by a factor of three. Taking into account this decay results in a ZPL weight of 0.63 at 5 K. If we furthermore assume that the upper two fine-structure states are dephasing fast due to phonon-assisted transitions to the lowest state, there is an additional decay by a factor of three. Taking all these corrections into account results in a ZPL weight of 0.91 at 5 K, as is shown in Figure 3d on the right *y*-axis. This weight is consistent with the ZPL weight 0.93 extracted from single NC PL (see SI). The overall temperature dependence of the corrected ZPL weight is similar to the one observed for epitaxial InGaAs QDs.[27] In CdSe colloidal quantum dots[19] instead, which have a much smaller exciton coherence volume, significantly stronger phonon assisted transition are present, with a ZPL weight of about 0.7 at 5 K, decaying to 0.4 at 20 K. However, the ZPL weight for temperatures below 12 K is found to be constant, in contrast to the two above mentioned systems, which both show a continuous decrease of the ZPL weight with temperature. Notably, the phonon-assisted transitions in these other QD systems are dominated by coupling to a continuum of acoustic phonons with linear dispersion around zero momentum (Γ-point), resulting in a broad phonon-assisted band. The single QD PL of the perovskite NCs (see SI) instead shows no such broad band, but rather two well-defined phonon energies of about 3 and 6 meV. These discrete phonon energies would be expected to lead to an activated behavior of the ZPL weight reduction, consistent with the observed constant ZPL weight below 12 K.

To further investigate the initial oscillations in the density dynamics in Figure 2a, we took data for different delays $\tau_{12} \geq 0$. The resulting FWM field amplitude as a function of $\tau_{13}$ is shown in Figure 4a. The amplitude of the oscillations changes while the oscillation period remains stable at $\tau_0 = 1.20 \pm 0.05$ ps. The relative amplitudes *B* of the damped oscillations are shown in Figure 4b as a function of time delay $\tau_{12}$, exhibiting an oscillating behavior. The red curve displays a fit with a damped squared-sine function. From this, an oscillation period of 1.22 ± 0.02 ps is obtained, which concurs with the period of the initial oscillations in the density decay. We therefore attribute these oscillations to coherent exciton-phonon coupling, resulting from the modulated polarization as a function of the harmonic nuclei displacement, as previously reported for CdSe[28,29] and PbS[30] NCs. The resulting vibrational energy of 3.45 ± 0.14 meV is in good agreement with the measured phonon energies of phonon-assisted transitions in single NC PL measurements (see SI), which have been attributed to TO-phonon modes in bulk $CsPbCl_3$[31] and for individual $CsPbBr_3$ NCs.[22] The oscillation damping could be due to decay of the phonon mode decay into acoustic phonons[29] and inhomogeneous broadening of the phonon mode in the QD ensemble, as observed in $FAPbBr_3$ NCs.[32] The three-pulse photon-echo signal $S(\tau_{12}, \tau_{13})$ in the direction $\mathbf{k}_S = -\mathbf{k}_1 + \mathbf{k}_2 + \mathbf{k}_3$ of an inhomogeneously broadened two-level system coupled to a single harmonic mode of angular frequency $\omega$ can be calculated as:[29,33]

$$S(\tau_{12}, \tau_{13}) \sim \exp[-4\Delta^2 (n(\omega) + 1) \cdot (1 - \cos(\omega\tau_{12}))(1 - \cos(\omega\tau_{13}))] \cdot \exp\left(-\frac{4\tau_{12}}{T_2}\right)$$

$$\cdot \exp\left(-\frac{2\tau_{13}}{T_1}\right) \cdot \Phi(\tau_{12}, \tau_{13}) \qquad (1)$$

Here, $\omega$ is the mode frequency and $n(\omega) = 1/[\exp\left(\frac{\hbar\omega}{k_B T}\right) - 1]$ is the Bose occupation factor. The function $\Phi(\tau_{12}, \tau_{13}) = 1 + \cos(\Delta^2[\sin(\omega\tau_{12}) - \sin(\omega\tau_{13}) + \sin(\omega\tau_{13} - \omega\tau_{12})])$ arises from the superposition of different Liouville space pathways.[34] $\Delta$ represents the coupling factor between excitonic state and phonon mode, and can be expressed by the Huang-Rhys parameter. For further details, we refer to the work of *Mittelmann* and *Schoenlein et al.*[28,29] In Figure 4c, we show the modelled electric fields of the three-pulse photon echo with $E(\tau_{12}, \tau_{23}) \sim \sqrt{S(\tau_{12}, \tau_{23})}$ according to equation (1). The model uses a Huang-Rhys parameter of $\frac{1}{2}\Delta^2 = 0.038$, a phonon mode at 3.45 meV (as measured in single NC PL measurements, see SI) and the above measured decay rates $\Gamma_1$ and $\gamma_{1,2}$ with their corresponding amplitudes. We have included an oscillation damping ($\tau_{damp} = 2.5\ ps$) and a finite pulse duration by multiplying $S(\tau_{12}, \tau_{13})$ with $\frac{1}{4}(1 + \text{erf}(\frac{\tau_{12}}{\tau_0}))(1 + \text{erf}(\frac{\tau_{13}}{\tau_0}))$, where $\tau_0 = 72$ fs describes the amplitude FWHM using the intensity FWHM $2\sqrt{\ln 2}\tau_0 = 120$ fs of the excitation pulses. The complete formula is provided in the SI. For better comparison with experiment, the calculated photon-echo signals for the experimental time delays are plotted in Figure 4d and Figure S5 in the SI, from which good agreement with measurements can be inferred.

In conclusion, we have investigated the coherence and density dynamics in fully inorganic CsPbBr$_2$Cl NCs at cryogenic temperatures. Using three-beam FWM, we obtain a dephasing time and a density decay time of several tens of picoseconds at 5 K. Furthermore, we find excitation of coherent phonons of 3.45 meV energy with a Huang-Rhys parameter of 0.038. The observed long dephasing time close to the lifetime limit is promising for applications in microcavity devices based on strong light-matter interaction.

AUTHOR INFORMATION

**Corresponding Authors (*)**

* [mvkovalenko@ethz.ch](mailto:mvkovalenko@ethz.ch) (M.V.K.)

* [langbeinww@cardiff.ac.uk](mailto:langbeinww@cardiff.ac.uk) (W.W.L.)

* [rfm@zurich.ibm.com](mailto:rfm@zurich.ibm.com) (R.F.M.)

**Author Contributions**

The manuscript was written through contributions of all authors. All authors have given approval to the final version of the manuscript.

**Notes**

The authors declare no competing financial interest.

ACKNOWLEDGMENT

The authors acknowledge helpful discussions with D.J. Norris. M.A.B., G.R., M.V.K., T.S., and R.F.M. acknowledge the European Union's Horizon-2020 program through the Marie-Sklodowska Curie ITN network "PHONSI" (H2020-MSCA-ITN-642656) and the Swiss State Secretariat for Education Research and Innovation (SERI). L.S., F.M. and W.L. acknowledge partial funding from EPSRC by the grants EP/M020479/1 and EP/M012727/1.

FIGURES

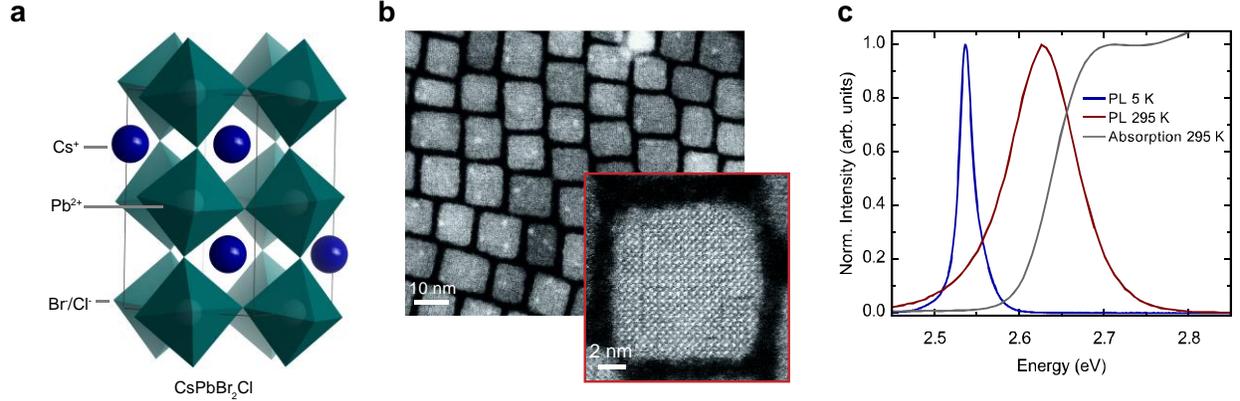

**Figure 1.** Crystal structure and optical properties of CsPbBr$_2$Cl nanocrystals. (a) Fully-inorganic perovskite cesium lead halide unit cell with typical quasi-cubic crystal structure with $\gamma$-orthorhombic distortion. (b) Transmission electron microscopy (TEM) image of CsPbBr$_2$Cl nanocrystals (inset: single NC with high-resolution TEM). (c) Normalized ensemble PL and absorption spectra at 295 K and 5 K.

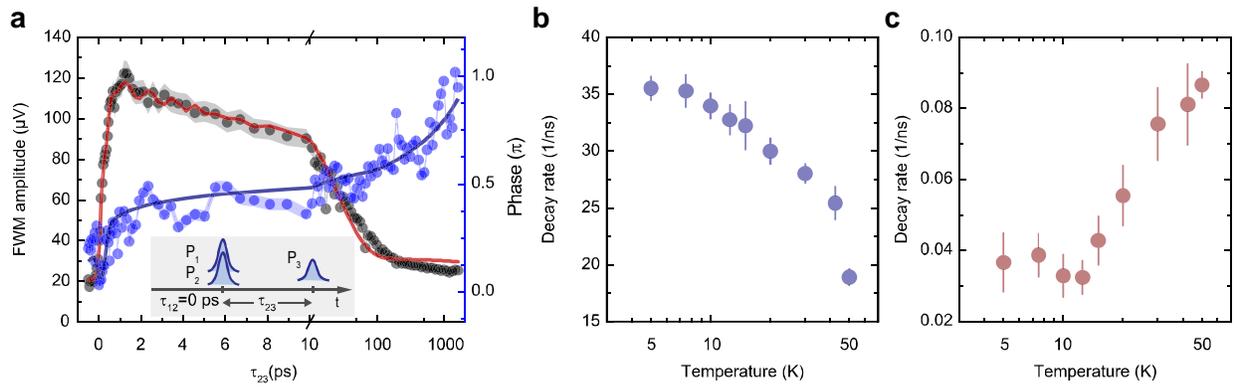

**Figure 2.** Temperature-dependent population density decay. (a) FWM field amplitude (black circles) and phase (blue circles) with their respective errors (shades) as a function of time delay $\tau_{23}$ between the first two pulses and the third pulse at 5 K. The solid lines represent fits to the data using the model discussed in the text. Inset: Sketch of the three-beam pulse sequence with $\tau_{12} = 0$ ps. (b) Decay rates $\Gamma_1$ and (c) $\Gamma_2$ of the two exponential decays extracted from the complex fit as function of temperature.

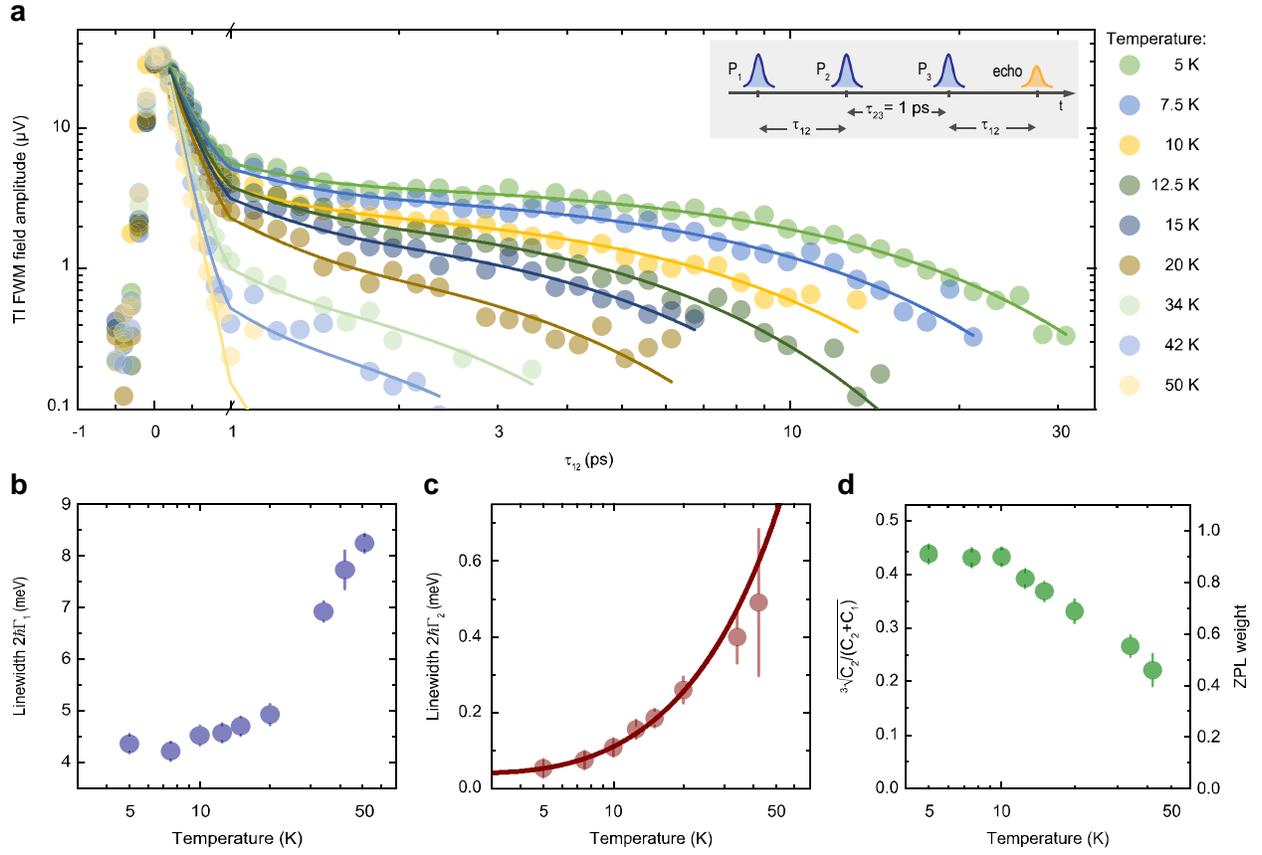

**Figure 3.** Temperature-dependent photon echo. (a) Time-integrated FWM field amplitude as a function of the time delay $\tau_{12}$ between the first and the second pulse in the temperatures range 5–51 K for fixed $\tau_{23} = 1$ ps. Inset: Sketch of the three-beam pulse sequence and resulting photon echo. (b) Linewidth $2\hbar\gamma_1$ as a function of temperature. (c) Homogeneous linewidth $2\hbar\gamma_2$ as a function of temperature. The line is a weighted fit to the data. (d) Temperature dependence of the nominal ZPL weight (left *y*-axis) and the corrected ZPL weight (right *y*-axis).

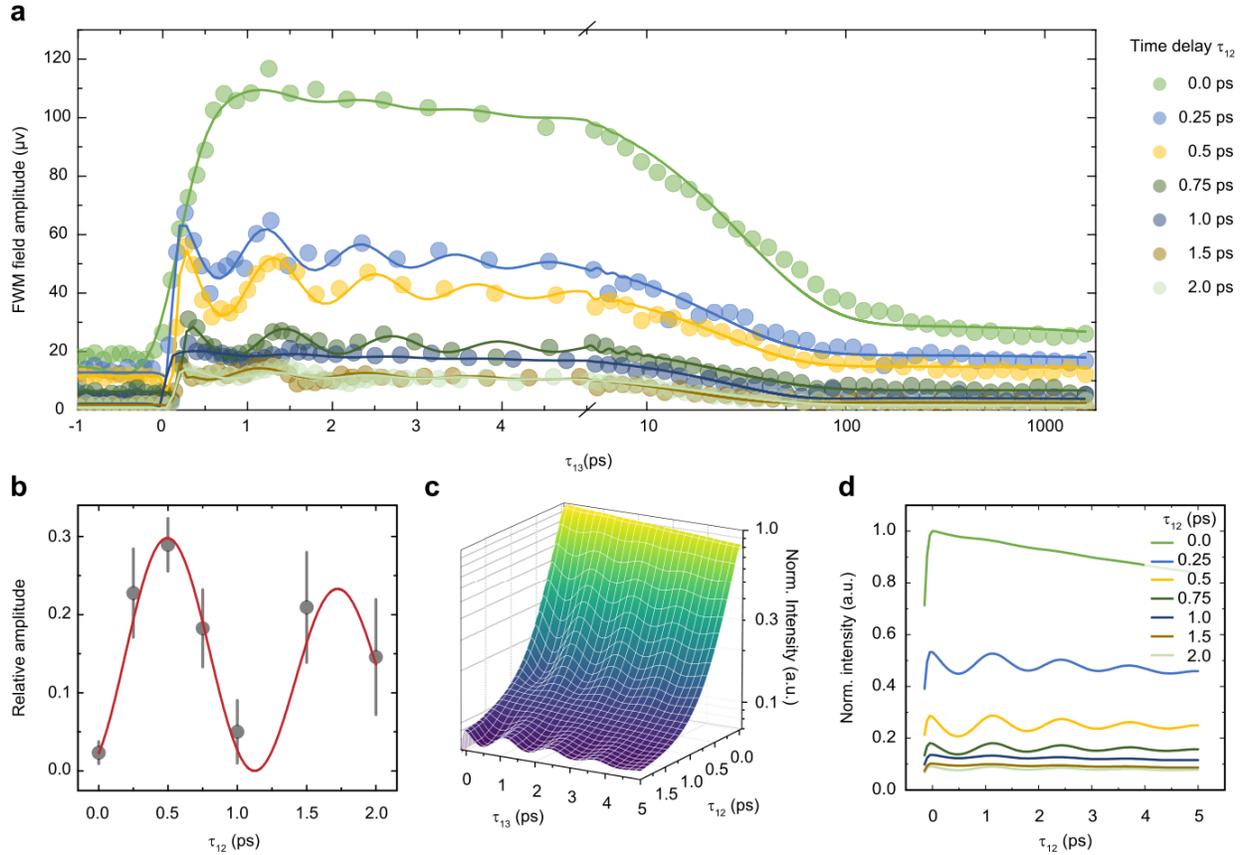

**Figure 4.** Coherent coupling to phonon modes. (a) Three-beam photon-echo signal as a function of $\tau_{13}$ for various time delays $\tau_{12}$ at 5 K. Fits to the data (circles) are represented by solid lines. (b) Relative amplitude of the damped oscillation from the fits in (a) as a function of the time delay $\tau_{12}$. The data are fitted with a damped sinusoidal function. (c) Calculated electric field of three-beam photon-echo signal for a two-level system coupled to a single harmonic mode as a function of $\tau_{12}$ and $\tau_{13}$ (linear scale). (d) Calculated three-beam photon-echo signal for a two-level system coupled to a single harmonic mode for similar time delays as in (a).

# Supplementary Information: Long Exciton Dephasing Time and Coherent Phonon Coupling in CsPbBr$_2$Cl Perovskite Nanocrystals

*Michael A. Becker[1,2], Lorenzo Scarpelli[3], Georgian Nedelcu[4,5], Gabriele Rainò[4,5], Francesco Masia[3], Paola Borri[3,6], Thilo Stöferle[1], Maksym V. Kovalenko[4,5]\*, Wolfgang Langbein[3]\*, Rainer F. Mahrt[1]\**

[1]IBM Research−Zurich, Säumerstrasse 4, 8803 Rüschlikon, Switzerland
[2]Optical Materials Engineering Laboratory, ETH Zürich, 8092 Zürich, Switzerland.
[3]School of Physics and Astronomy, Cardiff University, The Parade, Cardiff CF243AA, United Kingdom
[4]Institute of Inorganic Chemistry, Department of Chemistry and Applied Bioscience, ETH Zürich, 8093 Zürich, Switzerland.
[5]Laboratory of Thin Films and Photovoltaics, Empa – Swiss Federal Laboratories for Materials Science and Technology, 8600 Dübendorf, Switzerland.
[6]Cardiff University School of Biosciences, Museum Avenue, Cardiff CF10 3AX, United Kingdom

## Materials and Methods

**Chemicals.** The following reagents were used to prepare CsPbBr$_2$Cl nanocrystals: cesium carbonate (Cs$_2$CO$_3$, Aldrich, 99.9%), 1-octadecene (ODE, Sigma-Aldrich, 90%), oleic acid (OA, Sigma-Aldrich, 90%), oleylamine (OAm, Acros Organics, 80%–90%), lead chloride (PbCl$_2$, ABCR, 99.999%), lead bromide (PbBr$_2$, ABCR, 98%), n-trioctylphosphine (TOP, Strem, 97%), hexane (Sigma-Aldrich, ≥95%), toluene (Sigma-Aldrich, puriss. p.a., ACS reagent, ≥99.7%) and acetonitrile (Fisher Scientific, HPLC gradient grade).

**Synthesis and sample preparation**. The CsPbBr$_2$Cl NCs were prepared by following the procedure reported by Protesescu *et al.*,[1] with some adjustments. PbBr$_2$ (0.122 mmol) and PbCl$_2$ (0.0658 mmol) were loaded, inside of a glovebox, into a 3-neck flask along with pre-dried OA (1 mL), OAm (1 mL), TOP (1 mL) and ODE (5 mL). The flask was transferred to a Schlenk line and evacuated for 10 minutes at 120°C. The reaction mixture was heated up to 200°C under N$_2$ and 0.4 mL of hot Cs-oleate (prepared as described by Protesescu *et al.* [1]) was swiftly injected. The reaction was stopped after 10 seconds by immersing the flask into a water-bath. The solution was centrifuged (4 min, 13,750 g) and the supernatant discarded. Hexane (0.3 ml) was added to the precipitate to disperse the NCs, and the mixture was then centrifuged again. The supernatant was collected separately, and 0.9 mL toluene was added. The NCs were precipitated by adding 0.24 mL acetonitrile and centrifuged (3 min, 6,740 g). The obtained precipitate was re-dispersed in 2 mL toluene and filtrated. For the sample preparation, we added 5 m% polystyrene in toluene in a 1:2 ratio to the solution of nanocrystals and immediately drop-casted on *c*-cut quartz substrates. The film had a thickness of 17 μm, and showed a certain degree of agglomeration, as can be seen in Figure S1. However, scattering was not an issue in the alignment of the box geometry in three-pulse FWM (see SI: *Transient Four-Wave Mixing and Heterodyne Detection*).

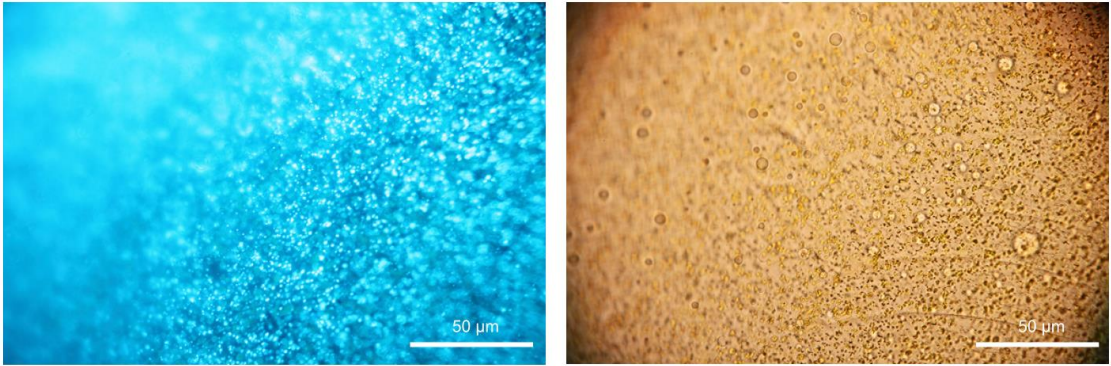

**Figure S1.** Fluorescence (left, excitation 410 – 440nm) and transmission (right, condenser with numerical aperture *NA*=0.8) microscope image using a 100× *NA*=1.3 oil immersion objective.

**Transient Four-Wave Mixing and Heterodyne Detection**

The homogeneous linewidth of the spectral absorption is inversely proportional to the dephasing time $T_2$. In an ensemble measurement, the absorption linewidth is inhomogeneously broadened, thus making it impossible to deduce the microscopic $T_2$ time from the absorption lineshape. Here, we make use of the third-order non-linearity of a material and perform four-wave mixing in the transient coherent domain after pulsed excitation to measure the $T_2$ time in presence of inhomogeneous broadening. For the heterodyne detection scheme, the pulse train of the excitation laser is divided into the excitation pulses and a reference pulse. In the transient degenerate three-beam FWM configuration, three excitation pulses resonant to the absorption of $CsPbBr_2Cl$ ($E_{Laser} = 2.6$ eV) with a repetition rate $\frac{\Omega_P}{2\pi} = 76.11$ MHz are used. To distinguish between the different non-linear orders, a so-called box geometry of the three excitation beams is used. According to the phase-matching conditions (i.e., momentum conservation), the third-order nonlinear signal is emitted in the direction $\boldsymbol{k}_S = \boldsymbol{k}_3 + \boldsymbol{k}_2 - \boldsymbol{k}_1$, and is spatially selected by means of an iris. The signal is detected in a balanced heterodyne detection scheme, where the signal interferes with a reference beam at a beam-splitter, and the intensities of the two resulting beams are measured with two photodiodes. The photo-current is proportional to the square of the interfering incoming complex fields of the reference and signal pulse train. Furthermore, a frequency selection scheme allows to discriminate the order of the non-linear polarization. Hereby, the optical frequencies are slightly shifted by radio-frequency amounts $\Omega_j$ using acousto-optic modulators. Using pulse trains that exhibit controlled phase variations given by $e^{-i\Omega_j t}$, the FWM signal can be detected at the frequency $\Omega_d = \Omega_3 + \Omega_2 - \Omega_1 - \Omega_P$ using lock-in amplifiers. This frequency selection scheme together with the interferometric detection of the FWM field amplitude constitutes the heterodyne detection scheme. For further information about the heterodyne detection scheme and the optical setup, we refer to refs. [21,35].

**Single CsPbBr$_2$Cl NC spectroscopy: Fine-structure splitting and phonon replica.**

Knowing the emission properties of single NCs helps to interpret the results of the FWM experiments on ensembles of CsPbBr$_2$Cl NCs. Single NC spectroscopy of lead halide perovskite nanocrystals reveals a bright triplet emission with three orthogonal fine-structure states. On average, the fine-structure splitting for three emission peaks is $\Delta_{FSS} = 1.15 \pm 0.26$ meV with a large distribution ranging from several hundred $\mu$eV up to meV.[5] In Figure S1, a spectrum of a single CsPbBr$_2$Cl NC at 5 K is shown (see ref. 5 for details on the experiment) that exhibits two emission peaks with a fine-structure splitting of 2 meV. Additionally, we observe two phonon replica, red-shifted by 3.1 and 6.6 meV with Huang-Rhys factors of 0.038 and 0.033, respectively. In literature, these phonon replica were attributed to TO-phonon replica.[22]

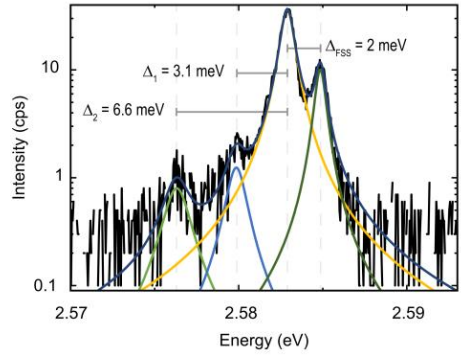

**Figure S2.** Single CsPbBr$_2$Cl NC PL measurement. The nanocrystal exhibits a fine structure with two emission peaks split by $\Delta = 2.0$ meV. Furthermore, the phonon replica at $\Delta_1 = 3.1$ meV and $\Delta_2 = 6.6$ meV are observed with Huang-Rhys factors of $S_1 = 0.038$ and $S_2 = 0.033$, respectively, resulting in a ZPL weight of 0.93.

**Heterodyne-Detected Four-Wave Mixing: Complex Fit to the Exciton Population Dynamics**

For a quantitative analysis, the population decay data is fitted with the complex multi-exponential response function, as explained in detail in ref. [35]:

$$R(\tau) \sim A_{nr} e^{i\phi_{nr}} \delta(\tau) + \sum_j A_j \Theta(\tau) \cdot e^{\left(i\phi_j - \frac{\tau}{\tau_j}\right)}.$$

Here, $A_j$, $\phi_j$ and $\tau_j$ are amplitude, phase and decay time of the *n*-th decay process. $A_{nr}$ is a non-resonant instantaneous component to account for effects like two-photon absorption and Kerr effect. The above equation describes the FWM as a coherent superposition of exponential decay components with their respective phases, given by their relative effect on absorption and refractive index. Slow drifts of the setup, for example due to room temperature changes, can affect the relative phase of the reference and probe pulses over a timescale of minutes. The fitting procedure accounts for this phase drift with a prefactor $e^{-i(\phi_0+\phi_0' t)}$, corresponding to a linear time-dependence of the phase, where $t$ is the real time during the measurement. The excitation pulse is taken into account by convoluting the response with a periodic Gaussian of full-width half-maximum $2\sqrt{\ln(2)}\,\tau_0$ in amplitude, given by the laser pulse auto-correlation. Furthermore, we include a pile-up of the signal $\sim \left(e^{\frac{T_R}{\tau_j}}-1\right)^{-1}$ due to the finite repetition rate $T_R$ of the excitation pulses of 13 ns, relevant for decay components with lifetimes similar or longer than $T_R$. Note that for lifetimes similar or longer than the modulation period of $\frac{1}{\Omega_1-\Omega_2}=160$ ns, the pilep-up would be saturated; a regime which is not considered explicitly in the fit formula as it is not relevant to the presented fits. In the measurements, initial oscillations in the density decay can be seen. We model this by multiplying the decay components, excluding pile-up, by $(1+Be^{-\gamma_p \tau}\cdot\cos(\omega_P\tau))$ with amplitude $B$, damping rate $\gamma_p$ and angular frequency of the oscillations $\omega_p$. The fit function of the density decay uses two exponential decay processes, yielding

$$F(t,\tau) = e^{-i(\phi_0+\phi_0' t)}\left[A_{nr}e^{i\phi_{nr}-\frac{\tau^2}{\tau_0^2}} + \sum_{j=1,2} A_j \left(\frac{1}{e^{\frac{T_R}{\tau_j}}-1} + \frac{1+Be^{-\gamma_p\tau}\cdot\cos(\omega_p\tau)}{2}\right)\left(1+\mathrm{erf}\left(\frac{\tau}{\tau_0}-\frac{\tau_0}{2\tau_j}\right)\right)\cdot e^{\left(i\phi_j+\frac{\tau_0^2}{4\tau_j^2}-\frac{\tau}{\tau_j}\right)}\right].$$

Furthermore, we measure the excitation-power dependent density grating decay at 5 K, shown in Figure S3. Again, we fit the data with the above described complex fit. The resulting decay times are independent of the excitation density within error. From the complex fits, we obtain an average initial decay time $\tau_1 = 35.6 \pm 1.5$ ps.

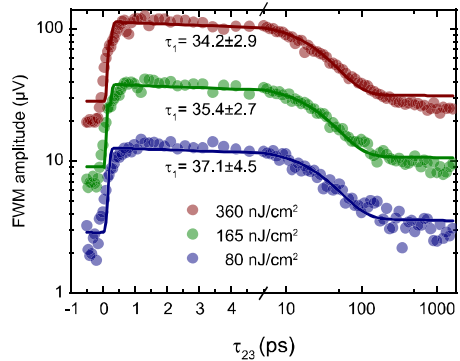

**Figure S3.** Excitation-power dependent FWM Field amplitude.

In Figure 2 in the main text, the temperature dependence of the first and the second decay component is shown. In Figure S4, we additionally plot the relative amplitude of the first decay component as a function of temperature, which stays almost constant within the temperature range measured.

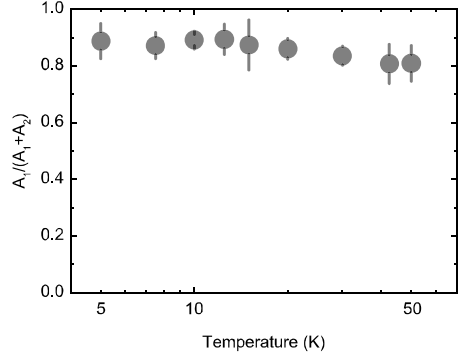

**Figure S4.** Relative amplitude of the first decay component of the bi-exponential complex fit as a function of temperature.

One can also use three exponential decay processes, resulting in a better fit at intermediate delay times around 100 ps. However, this leads to two time-constants in the 11 ps and 62 ps range, and is evidence for an inhomogeneous distribution of decay constants in the sample. We have therefore opted to use only two time-constants to extract the average decay, consistent with the analysis of the exciton dephasing.

**Coherent exciton-phonon coupling: Measurement results and theoretical model.**

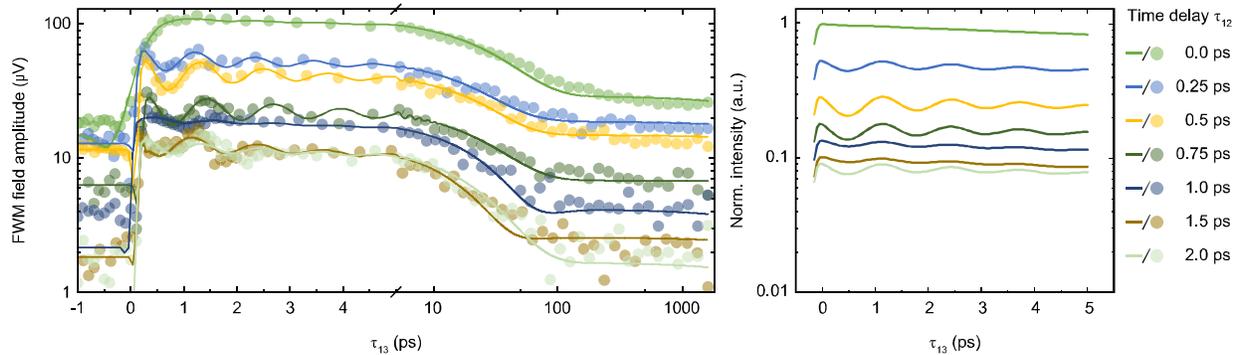

**Figure S5.** *Left panel:* Three-beam photon-echo signal as a function of $\tau_{13}$ for various time delays $\tau_{12}$ at 5 K (left). Fits to the data (circles) are represented by solid lines. *Right panel*: Calculated three-beam photon-echo signal for a two-level system coupled to a single harmonic mode for similar time delays as in the left panel.

For a better comparison, we plot the data of Figure 4a and d again in semi-log scale next to each other in Figure S5. As discussed in the main text, we use the model proposed by *Mittelmann* and *Schoenlein et al.*[28,29] to calculate the electric fields $E(\tau_{12}, \tau_{23}) \sim \sqrt{S(\tau_{12}, \tau_{23})}$ of the three-pulse photon-echo measurement in the presence of a single phonon mode. We added a damping term $\exp\left(-\frac{\tau_{12}}{\tau_{Damp}}\right) \exp\left(-\frac{\tau_{13}}{\tau_{Damp}}\right)$ to the oscillations, and a rise-term $\frac{1}{4}\left(1 + \text{erf}\left(\frac{\tau_{12}}{\tau_0}\right)\right)\left(1 + \text{erf}\left(\frac{\tau_{13}}{\tau_0}\right)\right)$ with a finite excitation pulse duration to the initial formula. Furthermore, we used a bi-exponential decay $A_1 \exp\left(-\frac{2\tau_{12}}{T_2^1}\right) + A_2 \exp\left(-\frac{2\tau_{12}}{T_2^2}\right)$ for the $\tau_{12}$ dependency:

$$S(\tau_{12}, \tau_{13}) = \exp\left[-4\Delta^2(n(\omega) + 1) \cdot (1 - \cos(\omega\tau_{12})) \exp\left(-\frac{\tau_{12}}{\tau_{Damp}}\right)\right.$$
$$\left. \cdot (1 - \cos(\omega\tau_{13})) \exp\left(-\frac{\tau_{13}}{\tau_{Damp}}\right)\right] \cdot \left\{A_1 \exp\left(-\frac{2\tau_{12}}{T_2^1}\right) + A_2 \exp\left(-\frac{2\tau_{12}}{T_2^2}\right)\right\}^2$$
$$\cdot \exp\left(-\frac{2\tau_{13}}{T_1}\right) \cdot \Phi(\tau_{12}, \tau_{13}) \cdot \frac{1}{4}\left(1 + \text{erf}\left(\frac{\tau_{12}}{\tau_0}\right)\right)\left(1 + \text{erf}\left(\frac{\tau_{13}}{\tau_0}\right)\right)$$